\documentclass[aps,prd,twocolumn,groupedaddress,amsmath,amssymb]{revtex4}
\usepackage{graphicx}
\usepackage{dcolumn}
\usepackage{bm}
\usepackage{verbatim}
\usepackage{amsmath}
\usepackage{amsfonts}
\usepackage{amssymb}
\usepackage{graphicx}
\usepackage{bm}
\usepackage{color}
\usepackage{amscd}

\def\bea{\begin{eqnarray}}
\def\eea{\end{eqnarray}}

\begin{document}

\title{First Law of Inner Mechanics of Black Holes in Generalized Minimal Massive Gravity}

\author{Mohammad Reza Setare}
\email[]{rezakord@ipm.ir}
\affiliation{{\small {\em  Department of Science, Campus of Bijar, University of Kurdistan, Bijar, Iran.}}}
\author{Hamed Adami}
\email[]{hamed.adami@yahoo.com}
\affiliation{{\small{\em  Research Institute for Astronomy and Astrophysics of Maragha (RIAAM),\\
P.O. Box 55134-441, Maragha, Iran.}}}

\begin{abstract}
In this paper, we examine the validity of the first law of inner mechanics of black holes in Generalized Minimal Massive Gravity. We consider BTZ and spacelike warped black holes and show that the first law of inner mechanics is valid for given black holes in Generalized Minimal Massive Gravity. As we expect, due to the presence of the Lorentz Chern-Simons term in Lagrangian of considered model, the product of the entropies of the
inner and outer horizons depends on the mass as it happens in Topologically Massive Gravity.
\end{abstract}

\maketitle

\section{Introduction}\label{S.I}
It is well known that black holes have temperature and entropy, and obey the laws of thermodynamics \cite{1',2'}. One interesting question which we can ask is this: what is the underlying statistical mechanical source of the black hole thermal properties? According to the AdS/CFT correspondence \cite{3'} black holes are dual to quantum states in the dual CFT, however this correspondence does not give a clear description about the black hole interior. It seems that the CFT does not contain operators that can describe the interior of black holes. But Papadodimas and Raju \cite{4'} have introduced such operators. The interesting and important result of \cite{4'} is this: "if one is allowed to use different operators to describe the
interior of the black hole in different states of the CFT, gravity|then there is no need for
firewalls at the horizon" \cite{5',6',7'}.\\
In this paper we show that one can consider the inner thermodynamics of black holes similar to the outer-horizon thermodynamics. We investigate the validity of the first law of inner mechanics of black holes in Generalized Minimal Massive Gravity (GMMG) \cite{5} This model is realized
by adding the Chern-Simons deformation term, the higher derivative deformation term, and an extra term
to pure Einstein gravity with a negative cosmological constant. In \cite{5} it is discussed that this theory is free of negative-energy bulk modes, and also avoids the aforementioned ``bulk-boundary unitarity clash''. By a Hamiltonian analysis one can show that the GMMG model has
no Boulware-Deser ghosts and this model propagate only two physical modes. The validity of the first law of black hole inner mechanics in the framework of Topological Massive Gravity (TMG) \cite{1} has been studied in \cite{24}. Castro and Rodriguez \cite{8'} have obtained formula for the products of horizon areas for solutions with non-spherical horizon topology in 5 dimensions, and have found that it is independent of the mass. However it has been shown that the product of the areas of the inner and outer horizons is dependent to the mass in the framework of TMG \cite{24}. Here also we show that, due to the presence of the Lorentz Chern-Simons term in Lagrangian of GMMG model, the product of the entropies of the
inner and outer horizons depends on the mass.
\section{Conserved Charges in Chern-Simons-Like Theories of Gravity}\label{S.II}
 Chern-Simons-like theories of gravity (CSLTG) \cite{6} is a class of gravitational theories in $(2 + 1)$ dimensions which can be formulated in terms of dreibein, spin-connection and some auxiliary 1-form fields. Topological massive gravity, New massive gravity (NMG) \cite{2}, Minimal massive gravity (MMG) \cite{3}, Zewi-dreibein \cite{4} and GMMG are examples of  Chern-Simons-like theories of gravity .\\
Let $e^{a} = e^{a}_{\hspace{1.5 mm} \mu} dx^{\mu}$ be a Lorentz vector-valued 1-form, where $e^{a}_{\hspace{1.5 mm} \mu}$ denotes the dreibein. We use the lower case Greek letters for the spacetime indices, and the internal Lorentz indices are denoted by the lower case Latin letters. The metric signature is mostly plus. Spacetime metric $g_{\mu \nu}$ and dreibein are related as $g_{\mu \nu}=\eta_{ab} e^{a}_{\hspace{1.5 mm} \mu} e^{a}_{\hspace{1.5 mm} \nu}$, where $\eta_{ab}$ is Minkowski metric. The spacetime metric is invariant under a Lorentz gauge transformations (or equivalently local Lorentz transformations, see Appendix J of \cite{7}) $e^{a} \rightarrow \Lambda ^{a}_{\hspace{1.5 mm} b} e^{b}$, where $\Lambda \in SO(2,1)$. In 3D, one can define the dualized spin-connection 1-form as $\omega^{a}=\frac{1}{2}\varepsilon^{abc} \omega_{bc}$, where $\omega^{a}_{\hspace{1.5 mm} b}= \omega^{a}_{\hspace{1.5 mm} b \mu} dx^{\mu}$ is spin-connection 1-form. Likewise, dualized curvature and torsion 2-form can be defined as $R(\omega) = d \omega+\omega \times \omega$ and $T(\omega) = D(\omega)e=de+\omega \times e$, respectively. We use a 3D-vector algebra notation for Lorentz vectors in which contractions with $\eta _{ab}$ and $\varepsilon _{abc}$ are denoted by dots and crosses, respectively. Here, $D(\omega)$ denotes covariant exterior derivative with respect to dualized spin-connection.\\
Suppose $ a^{ra}=a^{ra}_{\hspace{3 mm} \mu} dx^{\mu} $ is the Lorentz vector valued 1-form, where $r=1, ..., N$ refers to the flavour index. In fact, $ a^{ra}$ is a collection of the dreibein, the dualized spin-connection and the auxiliary fields, i.e. $a^{ra}=\{ e^{a},\omega ^{a}, h^{a}, f^{a} , \cdots \}$, where $h^{a}$ and $f^{a}$ are Lorentz vector valued auxiliary 1-form fields. The CSLTG Lagrangian 3-form is given by
\begin{equation}\label{1}
 L= \frac{1}{2} \textbf{g}_{rs} a^{r} \cdot d a^{s}+ \frac{1}{6} \textbf{f}_{rst} a^{r} \cdot a^{s} \times a^{t},
\end{equation}
where $\textbf{g}_{rs} $ is a symmetric constant metric on the flavour space and $\textbf{f}_{rst}$ is the totally symmetric flavour tensor interpreted as the coupling constant.\\
Conserved charge perturbation conjugate to a vector field $\xi$ can be defined as \cite{8,9,11,14}
\begin{equation}\label{2}
  \delta Q (\xi) = \frac{1}{8 \pi G}\int_{\Sigma}\left( \textbf{g}_{rs} i_{\xi} a^{r} - \textbf{g} _{\omega s} \chi _{\xi} \right) \cdot \delta a^{s},
\end{equation}
where $\Sigma$ is a space-like codimension-2 surface and $\chi_{\xi}^{a}= \frac{1}{2} \varepsilon^{a}_{\hspace{1.5 mm} bc} \lambda^{ab}_{\xi}$, in which $\lambda^{ab}_{\xi}$ is generator of a Lorentz gauge transformation. Here, $i_{\xi}$ denotes interior product in $\xi$. To find conserved charges, one can take an integration from Eq.\eqref{2} over one-parameter path on the solution space \cite{8}. In this way, background contribution is subtracted and then we will find a finite amount for charge. The conserved charge defined by Eq.\eqref{2} is not only conserved for the Killing vectors which are admitted by spacetime everywhere, but also it is conserved for the asymptotic Killing vectors. In general, $\lambda_{\xi} ^{ab}$ is a function of spacetime coordinates and of the diffeomorphism generator $\xi$. To use Eq.\eqref{2} we have to find an expression for $\lambda_{\xi}^{ab}$ in terms of dynamical fields (i.e. we need a gauge fixing) when we compute conserved charges. An appropriate gauge fixing is \cite{9,11}:
\begin{equation}\label{3}
  \chi _{\xi} ^{a} = i_{\xi} \Omega ^{a} + \frac{1}{2} \varepsilon ^{a}_{\hspace{1.5 mm} bc} e^{b \mu} e^{c \nu} \nabla _{\mu} \xi _{\nu},
\end{equation}
where
\begin{equation}\label{4}
  \Omega^{a}= \frac{1}{2}e^{a}_{\hspace{1.5 mm} \alpha}\epsilon^{\alpha}_{\hspace{1.5 mm} \nu \beta}  e^{\beta}_{\hspace{1.5 mm} c} \nabla_{\mu} e^{c\nu} d x^{\mu},
\end{equation}
is dualized torsion-free spin-connection and $\nabla_{\mu}$ is covariant derivative with respect to the metric connection. Stationary black hole spacetime admits two Killing vectors $\partial_{t}$ and $\partial_{\phi}$. The conserved charges conjugate to the Killing vectors $\partial_{t}$ and $-\partial_{\phi}$ refer to mass, $\mathcal{M}=Q(\partial_{t})$, and angular momentum, $\mathcal{J}=Q(-\partial_{\phi})$, respectively.\\
Black hole entropy is conserved charge conjugate to the horizon-generating Killing vector field \cite{15}. The horizon-generating Killing vector field vanishes on the bifurcation surface. Using this fact, we can obtain a formula for black hole entropy \cite{8,9}:
\begin{equation}\label{5}
  \mathcal{S}=  \frac{1}{4G} \int_{\mathcal{B}} \frac{d \phi}{\sqrt{g_{\phi \phi}}} \textbf{g} _{\omega r} a^{r}_{\phi \phi},
\end{equation}
where $\mathcal{B}$ denotes the bifurcation surface.
\section{Generalized Minimal Massive Gravity}\label{S.III}
Generalized Minimal Massive Gravity can be described by four flavours of 1-form, $a^{r}= \{ e, \omega , h, f \}$ and the non-zero components of the flavour metric and the flavour tensor are
\begin{equation}\label{6}
\begin{split}
     & \textbf{g}_{e \omega}=\textbf{f}_{e \omega \omega}=-\sigma, \hspace{0.4 cm} \textbf{g}_{\omega f}=\textbf{f}_{f \omega \omega}=-\frac{1}{m^{2}}, \\
     & \textbf{g}_{e h}=\textbf{f}_{e h \omega}=1, \hspace{0.4 cm} \textbf{g}_{\omega \omega}=\textbf{f}_{\omega \omega \omega}=\frac{1}{\mu},\\
     & \textbf{f} _{eff}= -\frac{1}{m^{2}}, \hspace{0.4 cm} \textbf{f}_{eee}=\Lambda_{0},\hspace{0.4 cm} \textbf{f}_{ehh}= \alpha,
\end{split}
\end{equation}
where $\sigma$, $\Lambda _{0}$, $\mu$, $m$ and $\alpha$ are a sign, cosmological parameter with dimension of mass squared, mass parameter of the Lorentz Chern-Simons term, mass parameter of New massive gravity term and a dimensionless parameter, respectively.
\subsection{BTZ Black Hole}\label{S.III.A}
 Ba\~nados,  Teitelboim and  Zanelli (BTZ) black hole spacetime \cite{16} can be described by the following dreibein
\begin{equation}\label{7}
 e^{0}= N_{\text{BTZ}} dt , \hspace{0.3 cm} e^{1}= \frac{dr}{N_{\text{BTZ}}}, \hspace{0.3 cm}  e^{2}= r \left( d \phi +N_{\text{BTZ}}^{\phi} dt \right),
\end{equation}
with
\begin{equation}\label{8}
  N_{\text{BTZ}}^{2}=\frac{(r^{2}-r_{+}^{2})(r^{2}-r_{-}^{2})}{l^{2}r^{2}}, \hspace{0.5 cm}  N_{\text{BTZ}}^{\phi}= -\frac{r_{+}r_{-}}{lr^{2}},
\end{equation}
where $r_{\pm}$ are outer/inner horizon radiuses and $l$ is AdS$_{3}$ radius. One can use following ansatz
\begin{equation}\label{9}
   \omega^{a}= \Omega^{a}-\alpha H e^{a},\hspace{0.3 cm} h^{a}= H e^{a}, \hspace{0.3 cm} f^{a}= F e^{a},
\end{equation}
to show that BTZ black hole is a solution of GMMG provided that
\begin{equation}\label{10}
  \begin{split}
       & \frac{\sigma}{ l^{2}} - \alpha (1 + \sigma \alpha ) H ^{2} + \Lambda _{0} - \frac{F^{2}}{ m^{2}}=0, \\
       & - \frac{1}{\mu l^{2}} + 2 (1 + \sigma \alpha ) H + \frac{2 \alpha}{m^{2}} F H + \frac{\alpha ^{2}}{\mu} H^{2}=0, \\
       & - F + \mu (1 + \sigma \alpha ) H + \frac{\mu \alpha}{m^{2}} FH=0,
  \end{split}
\end{equation}
where $F$ and $H$ are constant parameters \cite{8}. For BTZ black hole Eq.\eqref{2} and Eq.\eqref{5} can be simplified as
\begin{equation}\label{11}
\begin{split}
   \delta Q_{\text{BTZ}}(\xi) = \frac{1}{8 \pi G} \int _{\Sigma} \biggl\{ & - \left( \sigma + \frac{\alpha H}{\mu} + \frac{F}{m^{2}} \right)\\
    \times & \left[ (i_{\xi} \Omega - \chi _{\xi} ) \cdot \delta e + i_{\xi} e \cdot \delta \Omega \right)] \\
    + \frac{1}{\mu} [ (i_{\xi}& \Omega - \chi _{\xi} ) \cdot \delta \Omega   + \frac{1}{l^{2}} i_{\xi} e \cdot \delta e ] \biggr\},
\end{split}
\end{equation}
\begin{equation}\label{12}
  \mathcal{S}_{\text{BTZ}} = - \frac{1}{4G} \int_ \textbf{H}\frac{d\phi}{\sqrt{g_{\phi \phi}}} \left[  \left(\sigma + \frac{\alpha H}{\mu}+ \frac{F}{m^{2}} \right) g _{\phi \phi} - \frac{1}{\mu} \Omega _{\phi \phi} \right]  ,
\end{equation}
where \textbf{H} stand for the horizon.
\subsection{Warped black holes}\label{S.III.B}
Spacelike warped AdS$_{3}$ black hole (WBH) \cite{17} can be described by the following dreibein
\begin{equation}\label{13}
\begin{split}
     & e^{0}= l N_{\text{WBH}} d \hat{t} , \hspace{0.3 cm} e^{1}= \frac{l d\hat{r}}{4 R_{\text{WBH}} N_{\text{WBH}}}, \\
     & e^{2}= l R_{\text{WBH}} \left( d \hat{\phi} +N_{\text{WBH}}^{\hat{\phi}} d\hat{t} \right),
\end{split}
\end{equation}
with
\begin{equation}\label{14}
\begin{split}
    N_{\text{WBH}}^{2}= & \frac{\zeta^{2} \nu ^{2}(\hat{r}-\hat{r}_{+})(\hat{r}-\hat{r}_{-})}{4 R_{\text{WBH}}^{2}}, \\
     N^{\hat{\phi}}_{\text{WBH}}= & \frac{\left| \zeta \right| (\hat{r}+\nu \sqrt{\hat{r}_{+} \hat{r}_{-}})}{2 R_{\text{WBH}}^{2}},  \\
     R_{\text{WBH}}^{2} = & \frac{1}{4} \zeta ^{2} \hat{r} \left[ \left( 1 - \nu ^{2} \right) \hat{r} + \nu ^{2} \left( \hat{r}_{+} + \hat{r}_{-} \right) + 2 \nu \sqrt{\hat{r}_{+} \hat{r}_{-}} \right],
\end{split}
\end{equation}
where $\hat{r}_{\pm}$ are outer/inner horizon radiuses. Here $\zeta$ and $\nu$  are free parameters which allow us to keep contact with \cite{18,19,20,21}. The spacetime described by dreibein \eqref{13} admits $SL(2,\mathbb{R})\times U(1)$ as isometry group. Therefore, one can write a symmetric-two tensor as $S_{\mu \nu}=a_{1} g_{\mu \nu} +a_{2} J_{\mu} J_{\nu}$, where $J=\partial_{\hat{t}}$. It has been shown that the WBH (described by dreibein \eqref{13}) together with following ansatz
\begin{equation}\label{15}
  \begin{split}
     h^{a}_{\hspace{1.5 mm \mu}}= & H_{1} e^{a}_{\hspace{1.5 mm \mu}} +H_{2} J^{a}J_{\mu}, \\
     f^{a}_{\hspace{1.5 mm \mu}}= & F_{1} e^{a}_{\hspace{1.5 mm \mu}} +F_{2} J^{a}J_{\mu},
  \end{split}
\end{equation}
is a solution of GMMG provided that
\begin{equation}\label{16}
\begin{split}
     & \frac{\zeta^{2}}{4l^{2}} - \frac{1}{2} \alpha l \left| \zeta \right| H_{2} - \alpha ^{2} H_{1} \left( H_{1} + l^{2} H_{2} \right) \\
     & - \left( 2F_{1}+l^{2} F_{2} \right)=0 ,
\end{split}
\end{equation}
\begin{equation}\label{17}
  -\frac{\zeta^{2}}{l^{4}} \left( 1-\nu^{2} \right) + \frac{3 \alpha}{2l} \left| \zeta \right| H_{2} + \alpha ^{2} H_{1} H_{2} +F_{2}=0,
\end{equation}
\begin{equation}\label{18}
  \begin{split}
     &-\left( 1+ \alpha \sigma\right) \left( 2 H_{1} + l^{2} H_{2} \right) - \frac{l}{2 m^{2}} \left| \zeta \right| F_{2}  \\
       & - \frac{\alpha}{m^{2}} \left[ 2 H_{1}F_{1} + l^{2} \left( H_{1}F_{2}+H_{2}F_{1} \right) \right] \\
       & +\frac{1}{\mu} \left( 2F_{1}+l^{2} F_{2} \right)=0,
  \end{split}
\end{equation}
\begin{equation}\label{19}
\begin{split}
     & -\frac{1}{\mu} F_{2} + \left( 1+ \alpha \sigma\right) H_{2} + \frac{3}{2lm^{2}} \left| \zeta \right| F_{2}  \\
     & + \frac{\alpha}{m^{2}} \left( H_{1}F_{2}+H_{2}F_{1} \right) =0,
\end{split}
\end{equation}
\begin{equation}\label{20}
  \begin{split}
     &- \frac{\zeta^{2}}{4l^{2}} \sigma + \frac{1}{2} \left( 1+ \alpha \sigma\right) l \left| \zeta \right| H_{2}\\
      & + \alpha \left( 1+ \alpha \sigma\right) H_{1} \left( H_{1} + l^{2} H_{2} \right) \\
       & - \Lambda _{0} + \frac{1}{m^{2}} F_{1} \left( F_{1}+l^{2} F_{2} \right)=0,
  \end{split}
\end{equation}
\begin{equation}\label{21}
\begin{split}
     & \frac{\zeta^{2}}{l^{4}} \left( 1-\nu^{2} \right) \sigma -\frac{3}{2l} \left( 1+ \alpha \sigma\right) \left| \zeta \right| H_{2} \\
     & - \alpha \left( 1+ \alpha \sigma\right) H_{1}H_{2} - \frac{1}{m^{2}} F_{1} F_{2}=0,
\end{split}
\end{equation}
where $H_{1}$, $H_{2}$, $F_{1}$, $F_{2}$ are constant parameters and $J^{a} = e^{a}_{\hspace{1.5 mm \mu}} J^{\mu} $ \cite{22}. For warped black holes Eq.\eqref{2} and Eq.\eqref{5} can be simplified as
\begin{equation}\label{22}
  \begin{split}
      \delta Q_{\text{WBH}}(\xi) = \frac{1}{8\pi G} \int_{\Sigma}\biggl\{& - \left( \sigma + \frac{\alpha H_{1}}{\mu} + \frac{F_{1}}{m^{2}}\right)\\
       & \times [ i_{\xi} e \cdot \delta  \Omega + \left( i_{\xi} \Omega - \chi _{\xi}\right)\cdot \delta e ]  \\
        +\frac{1}{\mu} \left( i_{\xi} \Omega - \chi _{\xi}\right) \cdot \delta \Omega &+ \alpha H_{2} \left( \frac{\alpha H_{2}}{\mu} + \frac{2 F_{2}}{m^{2}}\right) i_{\xi} \mathfrak{J} \cdot \delta \mathfrak{J} \\
        +\biggl[ - \frac{\zeta^{2}}{\mu l^{2}} \biggl( \frac{3}{4} - \nu^{2} \biggr)& + l \left| \zeta \right| \left( \frac{\alpha H_{2}}{\mu} + \frac{F_{2}}{m^{2}}\right) \biggr] i_{\xi} e \cdot \delta e \\
        -\left( \frac{\alpha H_{2}}{\mu} + \frac{F_{2}}{m^{2}}\right)& \left[ i_{\xi} \mathfrak{J} \cdot \delta \Omega + \left( i_{\xi} \Omega - \chi _{\xi}\right)\cdot \delta \mathfrak{J} \right] \\
        +\biggl[ \frac{\zeta^{2}}{\mu l^{4}} \left( 1 - \nu^{2} \right)& - \frac{3\left| \zeta \right|}{2l} \left( \frac{\alpha H_{2}}{\mu} + \frac{F_{2}}{m^{2}}\right) \biggr]\\
        & \times \left( i_{\xi} \mathfrak{J} \cdot \delta e + i_{\xi}e \cdot \delta \mathfrak{J} \right) \biggr\},
  \end{split}
\end{equation}
\begin{equation}\label{23}
\begin{split}
    \mathcal{S}_{\text{WBH}}= \frac{1}{4G} \int_\textbf{H} \frac{d\phi}{\sqrt{g_{\phi \phi}}} \biggl\{ & -\left(\sigma + \frac{\alpha H_{1}}{\mu} + \frac{F_{1}}{m^{2}} \right) g_{\phi \phi}  \\
      + \frac{1}{\mu} \Omega_{\phi \phi} &- \left( \frac{\alpha H_{2}}{\mu} + \frac{F_{2}}{m^{2}} \right) J_{\phi} J_{\phi} \biggr\},
\end{split}
\end{equation}
where $ \mathfrak{J}^{a}_{\hspace{1.5 mm} \mu}=J^{a}J_{\mu} $ for simplicity.
\section{Inner Mechanics}\label{S.IV}
Now, we investigate properties of the inner horizons of BTZ and Warped black holes in GMMG. We show that a First Law proposed in \cite{23} is satisfied at the inner horizons of given black holes. To do this, we perform arguments in parallel with \cite{24}.
\subsection{BTZ black holes}\label{S.IV.A}
Consider BTZ metric
\begin{equation}\label{24}
  ds^{2}=- N_{\text{BTZ}}^{2} dt^{2}+ \frac{dr^{2}}{N_{\text{BTZ}}^{2}}+ r^{2} \left( d \phi +N_{\text{BTZ}}^{\phi} dt \right)^{2}.
\end{equation}
Using Eq.\eqref{11}, one can compute mass
\begin{equation}\label{25}
   \mathcal{M}_{\text{BTZ}} =- \frac{1}{8 G} \left[ \left(\sigma + \frac{\alpha H}{\mu} + \frac{F}{m^{2}}\right) \frac{r_{+}^{2}+r_{-}^{2} }{l^{2}} + \frac{2 r_{+} r_{-}}{\mu l^{3}} \right] ,
\end{equation}
and angular momentum
\begin{equation}\label{26}
    \mathcal{J}_{\text{BTZ}}= - \frac{1}{8 G} \left[ \left(\sigma + \frac{\alpha H}{\mu} + \frac{F}{m^{2}}\right) \frac{2 r_{+} r_{-}}{ l }  + \frac{r_{+}^{2}+r_{-}^{2} }{ \mu l^{2}} \right] ,
\end{equation}
of BTZ black hole in GMMG \cite{8}. This results are independent of choosing integration surface $\Sigma$. Angular velocity of outer horizon (event horizon) is
\begin{equation}\label{27}
  \Omega_{H^{+}}= -N_{\text{BTZ}}^{\phi}\big|_{r=r_{+}}=\frac{r_{-}}{l r_{+}}.
\end{equation}
Similarly, for inner horizon, we have
\begin{equation}\label{28}
  \Omega_{H^{-}}= -N_{\text{BTZ}}^{\phi}\big|_{r=r_{-}}=\frac{r_{+}}{l r_{-}}.
\end{equation}
The outer/inner horizon-generating Killing vector fields are given by $\zeta_{H^{\pm}}= \partial_{t}+ \Omega_{H^{\pm}} \partial_{\phi}$. Surface gravity can be computed by
\begin{equation}\label{29}
  \kappa_{H^{\pm}}= \sqrt{-\frac{1}{2} \nabla_{\mu}(\zeta_{H^{\pm}})_{\nu} \nabla^{\mu}(\zeta_{H^{\pm}})^{\nu} }\bigg|_{H^{\pm}}.
\end{equation}
Surface gravity and Hawking temperature are related as $T_{H}= \kappa_{H}/2 \pi$, then the Hawking temperatures of the inner and outer horizons are
\begin{equation}\label{30}
  T_{H^{\pm}}=\frac{r_{+}^{2}-r_{-}^{2}}{2 \pi l^{2} r_{\pm}}.
\end{equation}
 Since the outer/inner horizon-generating Killing vector fields are given by $\zeta_{H^{\pm}}$ vanish on outer/inner horizon then we are allowed to use Eq.\eqref{12} to obtain outer/inner horizon entropy. The $\phi$-$\phi$ components of metric and dualized spin-connection on outer/inner horizon are given by
\begin{equation}\label{31}
  g_{\phi \phi} \big|_{H^{\pm}}= r_{\pm}^{2} , \hspace{0.7 cm} \Omega_{\phi \phi} \big|_{H^{\pm}}= -\frac{r_{+} r_{-}}{l} .
\end{equation}
Therefore, outer/inner horizon entropies can be computed as
\begin{equation}\label{32}
   \mathcal{S}_{\text{BTZ}}^{\pm}=-\frac{\pi}{2 G} \left[ \left(\sigma + \frac{\alpha H}{\mu} + \frac{F}{m^{2}}\right) r_{\pm} + \frac{r_{\mp}}{\mu l} \right].
\end{equation}
These entropies can be written as
\begin{equation}\label{33}
   \mathcal{S}_{\text{BTZ}}^{\pm}=\frac{\pi^{2} l}{3} \left( c_{R}T_{R} \pm c_{L}T_{L}\right),
\end{equation}
where
\begin{equation}\label{34}
  c_{R/L}= -\frac{3l}{2G} \left( \sigma + \frac{\alpha H}{\mu}+ \frac{F}{m^{2}} \pm \frac{1}{\mu l}\right),
\end{equation}
are central charges in GMMG \cite{8} and
\begin{equation}\label{35}
  T_{R/L}= \frac{r_{+} \pm r_{-}}{2 \pi l^{2}},
\end{equation}
are right/left-moving temperatures. It can be easily checked that the outer/inner horizon entropies given in Eq.\eqref{32} obey the first laws of outer/inner mechanics
\begin{equation}\label{36}
  \delta \mathcal{M}_{\text{BTZ}}= \pm T_{H^{\pm}} \delta \mathcal{S}_{\text{BTZ}}^{\pm}+ \Omega_{H^{\pm}} \delta \mathcal{J}_{\text{BTZ}}.
\end{equation}
The product of the inner and outer horizon entropies is
\begin{equation}\label{37}
     \mathcal{S}_{\text{BTZ}}^{+} \mathcal{S}_{\text{BTZ}}^{-} = \frac{\pi^{4} l^{2}}{9} \left( c_{R}^{2} T_{R}^{2} - c_{L}^{2} T_{L}^{2}\right),
\end{equation}
From Eq.(65) in Ref.\cite{8}, one can deduce that right/left-moving energies in GMMG are given by
\begin{equation}\label{38}
  E_{R/L}= \frac{\pi^{2} l}{6} c_{R/L} T_{R/L}^{2}.
\end{equation}
Thus, the product of the inner and outer horizon entropies can be written as
\begin{equation}\label{39}
     \mathcal{S}_{\text{BTZ}}^{+} \mathcal{S}_{\text{BTZ}}^{-} = \frac{2 \pi^{2} l}{3} \left( c_{R} E_{R} - c_{L} E_{L}\right).
\end{equation}
Since mass and angular momentum of BTZ black hole are related to right and left-moving energies as
\begin{equation}\label{40}
  \mathcal{M}_{\text{BTZ}}= E_{R}+E_{L}, \hspace{0.7 cm} \mathcal{J}_{\text{BTZ}}=\frac{1}{l}( E_{R}-E_{L}),
\end{equation}
then
\begin{equation}\label{41}
\begin{split}
   \mathcal{S}_{\text{BTZ}}^{+} \mathcal{S}_{\text{BTZ}}^{-} =  -\frac{ \pi^{2} l^{2}}{G} \biggl[& \left(\sigma + \frac{\alpha H}{\mu} + \frac{F}{m^{2}}\right) l \mathcal{J}_{\text{BTZ}} \\
     & +\frac{\mathcal{M}_{\text{BTZ}}}{\mu l} \biggr].
\end{split}
\end{equation}
Therefore in GMMG the product of the inner and outer horizon entropies depends on the mass of the black hole. It does not depend on mass when we set $\mu \rightarrow \infty$.
\subsection{Warped black holes}\label{S.IV.B}
Now, we consider warped black hole metric
\begin{equation}\label{42}
\begin{split}
   ds^{2}= & - N_{\text{WBH}}^{2} d \hat{t}^{2}+ \frac{d\hat{r}^{2}}{4 R_{\text{WBH}}^{2} N_{\text{WBH}}^{2}} \\
     & + R_{\text{WBH}}^{2} \left( d \hat{\phi} +N_{\text{WBH}}^{\hat{\phi}} d\hat{t} \right)^{2}.
\end{split}
\end{equation}
Using Eq.\eqref{22}, one can compute mass
\begin{equation}\label{43}
   \mathcal{M}_{\text{WBH}} = \frac{\left| \zeta \right|^{3} \nu^{4} \hat{c}_{L}}{48} \left( \hat{r}_{+}+\hat{r}_{-}+2\nu \sqrt{\hat{r}_{+}\hat{r}_{-}} \right) ,
\end{equation}
and angular momentum
\begin{equation}\label{44}
\begin{split}
   \mathcal{J}_{\text{WBH}}= -\frac{\zeta^{4} \nu^{4}}{384} \biggl\{ & \hat{c}_{L} \left( \hat{r}_{+}+\hat{r}_{-}+2\nu \sqrt{\hat{r}_{+}\hat{r}_{-}} \right)^{2} \\
     &  - \hat{c}_{R} \left( \hat{r}_{+}-\hat{r}_{-}\right)^{2} \biggr\} ,
\end{split}
\end{equation}
of warped black hole in GMMG \cite{22}. Here, $\hat{c}_{R/L}$ are defined as
\begin{equation}\label{45}
\begin{split}
   \hat{c}_{R}=-\frac{3 l }{\left| \zeta \right| \nu^{2} G}\biggl\{ & \sigma + \frac{\alpha}{\mu} \left(H_{1}+l^{2}H_{2} \right)+\frac{1}{m^{2}} \left(F_{1}+l^{2}F_{2} \right) \\
     & -\frac{\left| \zeta \right|}{2\mu l} \left( 1-2\nu^{2}\right)\biggr\},
\end{split}
\end{equation}
\begin{equation}\label{46}
\begin{split}
    \hat{c}_{L}=-\frac{3 l}{\left| \zeta \right| \nu^{2} G}\biggl\{& \sigma + \frac{\alpha}{\mu} \left(H_{1}+l^{2}H_{2} \right)+\frac{1}{m^{2}} \left(F_{1}+l^{2}F_{2} \right) \\
     & -\frac{\left| \zeta \right|}{2\mu l}\biggr\}.
\end{split}
\end{equation}
In Ref. \cite{22}, it has been shown that GMMG and spacelike warped boundary conditions can be regarded dual to a 2d CFT with central charges given in \eqref{45} and \eqref{46}. Angular velocity of outer/inner horizon is given by
\begin{equation}\label{47}
  \hat{\Omega}_{H^{\pm}}= -N_{\text{WBH}}^{\hat{\phi}}\big|_{\hat{r}=\hat{r}_{\pm}}=-\frac{2}{\left| \zeta \right| \left( \hat{r}_{\pm} + \nu \sqrt{\hat{r}_{+} \hat{r}_{-}} \right)}.
\end{equation}
By using Eq.\eqref{29}, one can obtain surface gravities and consequently Hawking temperatures of outer and inner horizons as
\begin{equation}\label{48}
  \hat{T}_{H^{\pm}}=\frac{\left| \zeta \right| \nu^{2}\left( r_{+}-r_{-} \right)}{4 \pi \left( r_{\pm} + \nu \sqrt{r_{+} r_{-}} \right)}.
\end{equation}
For warped black hole, we have
\begin{equation}\label{49}
  \begin{split}
      g_{\hat{\phi} \hat{\phi}} \big| _{H^{\pm}} = & \frac{1}{4} l^{2} \zeta^{2} \left( \hat{r}_{\pm} + \nu \sqrt{\hat{r}_{+} \hat{r}_{-}}\right)^{2}, \\
      \Omega_{\hat{\phi} \hat{\phi}} \big| _{H^{\pm}}=& \mp \frac{1}{4} \zeta^{2} \nu^{2} \sqrt{g_{\hat{\phi} \hat{\phi}}} \big| _{H^{\pm}} \left( \hat{r}_{+}-\hat{r}_{-} \right) + \frac{\left| \zeta \right|}{2 l} g_{\hat{\phi} \hat{\phi}}\big| _{H^{\pm}}, \\
      \left( J_{\hat{\phi}} J_{\hat{\phi}} \right) \big| _{H^{\pm}}&= l^{2} g_{\hat{\phi} \hat{\phi}}\big| _{H^{\pm}}.
  \end{split}
\end{equation}
By substituting Eq.\eqref{49} into Eq.\eqref{23}, we obtain outer/inner horizon entropies
\begin{equation}\label{50}
\begin{split}
   \mathcal{S}_{\text{WBH}}^{\pm}= &- \frac{\pi l \left| \zeta \right|}{4G} \biggl\{ \biggl[ \sigma + \frac{\alpha}{\mu} \left(H_{1}+l^{2}H_{2} \right)\\
   &+\frac{1}{m^{2}} \left(F_{1}+l^{2}F_{2} \right)-\frac{\left| \zeta \right|}{2\mu l} \biggr] \left( \hat{r}_{\pm} + \nu \sqrt{\hat{r}_{+} \hat{r}_{-}} \right)\\
     & \pm \frac{\left| \zeta \right| \nu^{2}}{2 \mu l} \left( \hat{r}_{+}-\hat{r}_{-}\right) \biggr\}.
\end{split}
\end{equation}
By defining right and left-moving temperatures as
\begin{equation}\label{51}
\begin{split}
     \hat{T}_{R}=&\frac{\zeta^{2} \nu^{2}}{8 \pi l} \left( \hat{r}_{+}-\hat{r}_{-}\right), \\
    \hat{T}_{L}= & \frac{\zeta^{2} \nu^{2}}{8 \pi l} \left( \hat{r}_{+}+\hat{r}_{-}+2\nu \sqrt{\hat{r}_{+}\hat{r}_{-}} \right),
\end{split}
\end{equation}
respectively, we can write outer/inner horizon entropies in the following form
\begin{equation}\label{52}
   \mathcal{S}_{\text{WBH}}^{\pm}=\frac{\pi^{2} l}{3} \left( \hat{c}_{R}\hat{T}_{R} \pm \hat{c}_{L}\hat{T}_{L}\right),
\end{equation}
which is the same as Eq.\eqref{33} for the BTZ black hole. Mass, angular momentum and outer/inner horizon entropies of warped black hole satisfy the first laws of outer/inner mechanics
\begin{equation}\label{53}
  \delta \mathcal{M}_{\text{WBH}}= \pm \hat{T}_{H^{\pm}} \delta \mathcal{S}_{\text{WBH}}^{\pm}+ \hat{\Omega}_{H^{\pm}} \delta \mathcal{J}_{\text{WBH}}.
\end{equation}
Similar to Eq.\eqref{38}, from Eq.(79) in Ref.\cite{22}, we can define right/left-moving energies for warped black hole in GMMG
\begin{equation}\label{54}
  \hat{E}_{R/L}= \frac{\pi^{2} l}{6} \hat{c}_{R/L} \hat{T}_{R/L}^{2},
\end{equation}
then the product of the inner and outer horizon entropies can be written as
\begin{equation}\label{55}
     \mathcal{S}_{\text{WBH}}^{+} \mathcal{S}_{\text{WBH}}^{-} = \frac{2 \pi^{2} l}{3} \left( \hat{c}_{R} \hat{E}_{R} - \hat{c}_{L} \hat{E}_{L}\right).
\end{equation}
Right and left-moving energies are related to mass and angular momentum of warped black hole as
\begin{equation}\label{56}
\begin{split}
    \hat{E}_{R}= & \frac{1}{l} \mathcal{J}_{\text{WBH}} + \frac{6}{l \zeta^{2} \nu^{4} \hat{c}_{L}}\mathcal{M}_{\text{WBH}}, \\
      \hat{E}_{L}=& \frac{6}{l \zeta^{2} \nu^{4} \hat{c}_{L}}\mathcal{M}_{\text{WBH}},
\end{split}
\end{equation}
then we can write the inner and outer horizon entropies as
\begin{equation}\label{57}
   \mathcal{S}_{\text{WBH}}^{+} \mathcal{S}_{\text{WBH}}^{-} =  \frac{ 2 \pi^{2}}{3} \biggl[ \hat{c}_{R} \mathcal{J}_{\text{WBH}} + \frac{18 \mathcal{M}_{\text{WBH}}^{2}}{\mu G \zeta^{2} \nu^{4} \hat{c}_{L}} \biggr].
\end{equation}
Again in GMMG the product of the inner and outer horizon entropies of warped black hole depends on the mass of the black hole.
\section{Conclusion}
We have summarized the method of obtaining conserved charges and black hole entropy in CSLTG. We have considered the GMMG model, as an example of the CSLTG. The BTZ and spacelike warped black holes are solutions of the GMMG model. We have shown that, for both black holes, outer/inner horizon entropies can be written as $\mathcal{S}^{\pm}=\frac{\pi^{2} l}{3} \left( c_{R}T_{R} \pm c_{L}T_{L}\right)$, where $T_{R/L}$ are right/left-moving temperatures and $c_{R/L}$ are central charges. These entropies satisfy the first law of outer/inner mechanics of black holes $\delta \mathcal{M}= \pm T_{H^{\pm}} \delta \mathcal{S}^{\pm}+ \Omega_{H^{\pm}} \delta \mathcal{J}$. The product of the entropies of the inner and outer horizons can be written in terms of right and left-moving energies. We have shown that, as topologically massive gravity case \cite{24}, the product of the inner and outer horizons entropies depends on the mass. This dependence comes from the contribution of the Lorentz Chern-Simons term in the Lagrangian. All of the obtained results in this paper will be reduce to the results in TMG \cite{24} when we set $\sigma=-1$, $\alpha=0$ and $m^{2} \rightarrow \infty$.
\section{Acknowledgments}
The work of Hamed Adami has been financially supported by Research Institute for Astronomy Astrophysics of Maragha (RIAAM).

\end{document}